\documentclass[11pt,twoside]{article}
\usepackage{asp2010}

\resetcounters

\begin{document}

\title{Two Populations of Companions around White Dwarfs:  The Effect of Tides and Tidal Engulfment}
\author{J. Nordhaus
\affil{Dept. of Astrophysical Sciences, Princeton University, Princeton, NJ 08544}
}

\begin{abstract}
During post-main-sequence evolution, radial expansion of the primary star, accompanied by intense winds, can significantly alter the binary orbit via tidal dissipation and mass loss.  The fate of a given binary system is determined by the initial masses of the primary and companion, the initial orbit (taken to be circular), the Reimers mass-loss parameter, and the tidal prescription employed.  For a range of these parameters, we determine whether the orbit expands due to mass loss or decays due to tidal torques. Where a common envelope (CE) phase ensues, we estimate the final orbital separation based on the energy required to unbind the envelope. These calculations predict period gaps for planetary and brown dwarf companions to white dwarfs.  In particular, the lower end of the gap is the longest period at which companions survive their CE phase while the upper end of the gap is the shortest period at which a CE phase is avoided.  For binary systems with 1 $M_\odot$ progenitors, we predict no Jupiter-mass companions with periods $\lesssim$270 days.  For binary systems consisting of a 1 $M_\odot$ progenitor with a 10 Jupiter-mass companion, we predict a close, post-CE population with periods $\lesssim$0.1 days and a far population with periods $\gtrsim$380 days.  These results are consistent with the detection of a $\sim$50 $M_{\rm J}$ brown dwarf in a $\sim$0.08 day orbit around the white dwarf WD 0137-349 and the tentative detection of a $\sim$2 $M_{\rm J}$ planet in a $\sim$4 year orbit around the white dwarf GD66.
\end{abstract}

\section{Introduction}
For low-mass stars (initially $\lesssim$8~M$_\odot$), post-main sequence (post-MS) evolution is
characterized by expansion via giant phases accompanied by the onset
of mass-loss.  During the Asymptotic Giant Branch phase
(AGB), dust-driven winds expel the stellar envelope as the star transitions to a white dwarf (WD).  Before formation of
the WD remnant, the spherical outflows observed during the AGB phase
undergo a dramatic transition to the highly asymmetric and often
bipolar geometries seen in all post-AGB and young planetary nebulae
(PNe; \citealt{Sahai:1998ee}).  This transition is often accompanied by
high-speed, collimated outflows.  For recent reviews see
\citet{van-Winckel:2003pi} and
\citet{de-Marco:2009vl}.

A central hypothesis to explain shaping in post-AGB/PNe is that a
close companion is necessary to power and shape bipolarity.  This is
supported by observations of excess momenta in all post-AGB
outflows relative to what isotropic radiation pressure can provide
\citep{Bujarrabal:2001bs}.  Additionally, maser
observations show magnetic jet collimation in AGB and young
post-AGB stars \citep{Vlemmings:2006vl,Sabin:2007zp,Vlemmings:2008ad,Amiri:2010fk}.
Such collimation supports the binary hypothesis because it is
difficult, if not impossible, for single AGB stars to generate the large field strengths
needed to power the outflows \citep{Nordhaus:2007il,Nordhaus:2008fe}.  If a close
companion is present, strong interactions can transfer energy and
angular momentum from the companion to the primary or outflow.  In particular, if
the companion is engulfed in a common envelope (CE), rapid in-spiral
can cause significant shear inside the CE
\citep{Nordhaus:2006oq, Nordhaus:2008lr}.  Coupled with a strong
convective envelope, large-scale magnetic fields are amplified and are
sufficient to unbind the envelope and power the outflow
\citep{Nordhaus:2007il}.  The recent detection of a white dwarf with an orbiting $\sim$50 $M_{\rm J}$ brown dwarf in a $\sim$2 hour orbit demonstrates that low-mass companions can survive common envelope phases (CEP) \citep{Maxted:2006fj}.  The detection of a planetary companion ($M{\rm sin}i=3.2$ $M_{\rm J}$) around the extreme horizontal branch star V391 Pegasi in a $\sim$1.7 AU orbit ($\sim$3.2 year period; \citealt{Silvotti:2007fk}) and the tentative detection of a $\sim$2~$M_{\rm J}$ planet in a $\gtrsim$2.7~AU orbit ($\gtrsim$4~year period; \citealt{Mullally:2008fk}) around the white dwarf GD66 provide further motivation for studying post-MS orbital dynamics.  For more detail on the work summarized in this proceeding, we refer the reader to \cite{Nordhaus:2010lr}.

\begin{figure}
\begin{center}
\includegraphics[width=8.7cm,angle=0,clip=true]{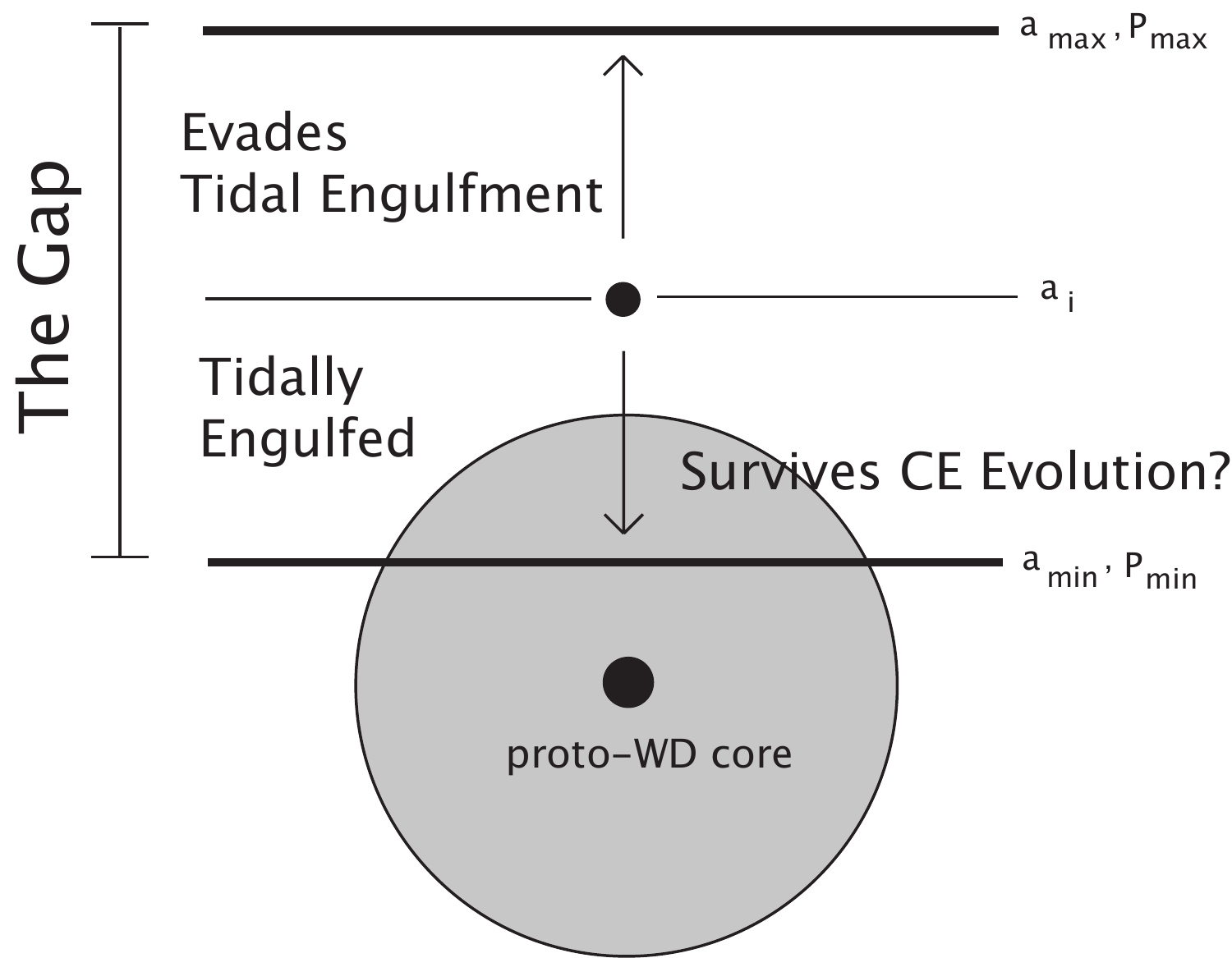}
\caption{The period gap for low-mass companions around white dwarfs.
  The orbit of a companion located initially at $a_i$ decays and
  plunges into the giant star.  Depending on the mass of the companion
  and stellar structure at plunge time, the companion may or may not
  survive the CE phase.  Companions slightly exterior to $a_i$ avoid
  engulfment and never enter a CE; their orbits expand due to
  mass-loss.  The gap is set by the final maximum semimajor axis which
  survives CE evolution ($a_{\rm min}$) and the final minimum
  semimajor axis which avoids tidal engulfment ($a_{\rm max}$).
\label{thegap}}
\end{center}
\end{figure}

\section{Tides and Mass-loss}

As the binary system evolves, mass-loss and tidal torques are in
competition.  Mass lost from the system acts to increase the semimajor
axis while tidal torques decrease it.  For each primary and companion,
we compute the evolution of the orbit from the zero-age main sequence
through the post-main sequence.  If the companion is tidally engulfed
(i.e.~plunges into the primary star), it enters a common envelope with
the primary.  If the companion evades tidal engulfment, mass-loss continues and the
orbit expands until the end of the evolutionary model.  We take the companion to be tidally locked to the primary, as is expected; in other contexts, this assumption might be testable \citep{Spiegel:2007zr}.  For a detailed description of the mass-loss prescriptions, tidal formalisms and orbital assumptions employed see \cite{Nordhaus:2010lr}.

For each stellar model, companion mass, and tidal theory, we calculate
the maximum initial semimajor axis, $a_{\rm i,max}$, that is tidally
engulfed (see Fig.~1).  Companions initially located exterior to $a_{\rm i,max}$ evade tidal engulfment and move outward while companions located interior to  $a_{\rm i,max}$ plunge into their host star.  Upon tidal engulfment, the companion enters a common envelope with the primary star \citep{Nordhaus:2006oq}.  The companion in-spirals until it is either tidally disrupted or supplies enough orbital energy to overcome the binding energy of the envelope and survive the CE phase.

\section{Period Gaps for Planets and Brown Dwarfs Around White Dwarfs}
\label{sec:PGaps}
We calculate the minimum period gap expected for a given binary system by assuming that all of the orbital energy released during in-spiral goes toward ejecting the CE.  This gives an upper bound on the inner orbital radius
at which we would expect to find companions that have survived a
CE phase (see Fig.~\ref{thegap}).  The outer orbital radius is the minimum position for which a companion evades tidal capture but migrates outward due to mass-loss from the primary.  These
calculations predict period gaps for planetary and brown dwarf companions to white dwarfs.

\begin{figure}
\begin{center}
\includegraphics[width=6.6cm,angle=0,clip=true]{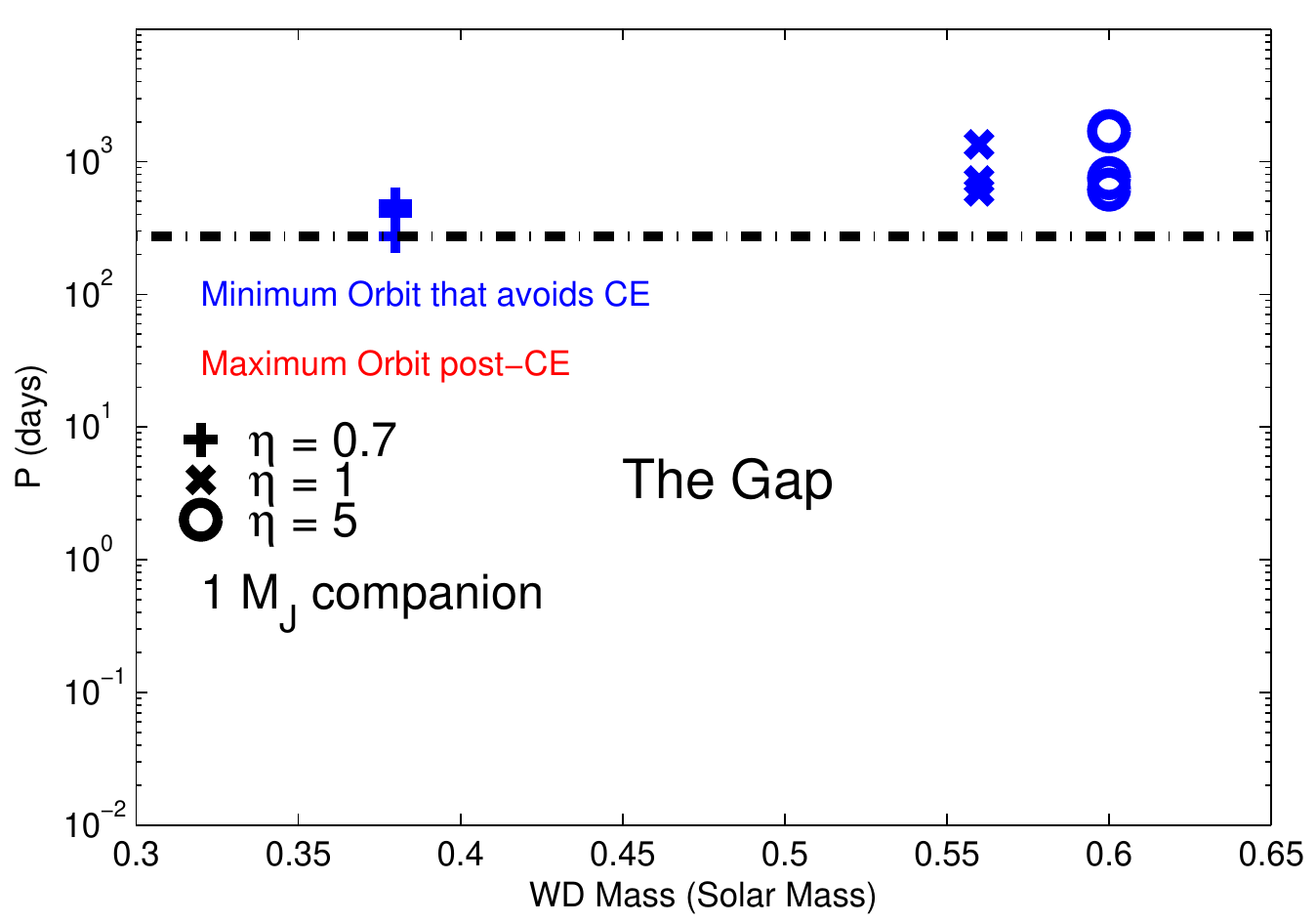}
\includegraphics[width=6.6cm,angle=0,clip=true]{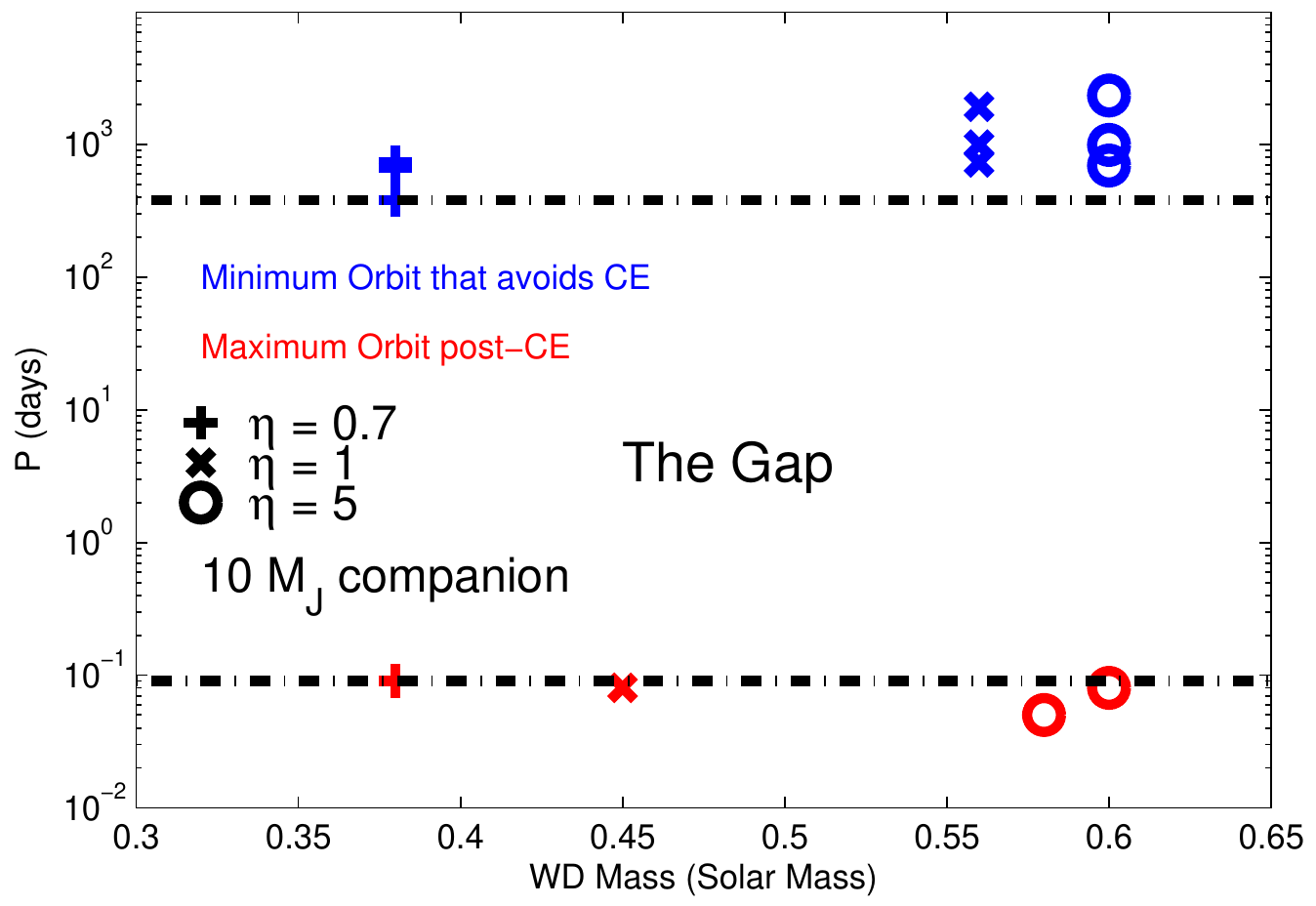}
\caption{The predicted period gaps for a 1 $M_\odot$ progenitor with 1
  $M_{\rm J}$ (left) and 10 $M_{\rm J}$ (right) companions.  The symbols represent
  different Remiers $\eta$ values for the various tidal prescriptions \citep{Nordhaus:2010lr}.  For the 1 $M_{\rm J}$ system, no companion
  survives CE evolution.  Thus, we predict a paucity of 1 $M_{\rm J}$
  companions with periods $\lesssim$270 days.  For the 10 $M_\odot$
  system, several companions survive CE evolution and are located in
  short-period orbits.  The predicted period gap occurs between
  $\sim$0.1 and 380 days.
\label{table_gap}}
\end{center}
\end{figure}

Our minimum period gaps are presented in Fig.~2.  We note that no Jupiter-mass companions survive CE evolution.  Thus, we predict a paucity of Jupiter-mass companions with periods $\lesssim$270~days around white dwarfs.  Additionally, our results predict that there should be a paucity of 10~$M_{\rm J}$ companions with periods between 0.1~days and 380~days.

\section{Conclusions}

By utilizing stellar evolution models from the ZAMS through the post-MS, we have followed the orbital dynamics of binary systems in which the companion is a planet or brown dwarf.  Dynamically, the orbital evolution is subject to mass-loss (which acts to increase the separation) and tidal torques (which act to decrease the separation).  For various tidal prescriptions and mass-loss rates, we determined the maximum separation for which companions might be tidally engulfed (i.e.~plunge into the primary star).  These results serve as initial conditions for the onset of the common envelope phase for low-mass companions.

For a binary system consisting of a 1~$M_\odot$ primary with a 1~$M_{\rm J}$ companion, we predict a paucity of Jupiter-mass companions with periods below $\sim$270~days.  For a 1~$M_\odot$ primary with a 10~$M_{\rm J}$ companion, the gap occurs between $\sim$0.1 and $\sim$380~days.  Note that our estimated gaps are conservative and are obtained by finding the minimum gap that might be expected for a range of mass-loss rates and a range of assumptions about tidal dissipation.  It is unlikely that the true gaps would be narrower than the ranges quoted above, but they easily could be wider.  As our knowledge of stellar evolution and tidal dissipation improves, so will our estimates of the ranges for these gaps.  Finally, we note that the results of surveys searching for low mass companions to white dwarfs might help to constrain theories of both stellar evolution and tides.

\bibliography{nordhaus}
\bibliographystyle{asp2010}

\end{document}